\documentclass[12pt]{article}%
\usepackage{amsfonts}
\usepackage{sw20bams}
\usepackage{amsmath}
\usepackage{amssymb}
\usepackage{graphicx}%
\providecommand{\U}[1]{\protect\rule{.1in}{.1in}}
\begin{document}

\title{Recursive PGFs for BSTs and DSTs}
\author{Steven Finch}
\date{February 7, 2020}
\maketitle

\begin{abstract}
We review fundamentals underlying binary search trees and digital search
trees, with (atypical) emphasis on recursive formulas for associated
probability generating functions. \ Other topics include higher moments of
BST\ search costs and combinatorics for a certain finite-key analog of DSTs.

\end{abstract}

\footnotetext{Copyright \copyright \ 2020 by Steven R. Finch. All rights
reserved.}Let $K$ denote the number of comparisons in a
successful/unsuccessful search of a random tree. \ Exact expressions for
probabilities/moments of $K$ exist for both binary search trees and digital
search trees \cite{Lou-tcs2, Mah-tcs2}. \ On the one hand, further exposition
on such well-known algorithms seems unnecessary. \ On the other hand, our
experience with recursively-defined probability generating functions in
\cite{Fi0-tcs2} illustrates the value of revisiting old topics with new
insights. \ We gather previously-scattered results into one place, trusting
that this work now will save some researchers the trouble later.

\section{Binary Search Trees}

Consider the R\ program:%
\[%
\begin{array}
[c]{l}%
\text{\texttt{f
$<$%
- function(x,V,k)}}\\
\text{\texttt{\{}}\\
\text{\texttt{\ \ if(NROW(V)==0) k
$<$%
- 0}}\\
\text{\texttt{\ \ else \{}}\\
\text{\texttt{\ \ \ \ u
$<$%
- V[1]}}\\
\text{\texttt{\ \ \ \ if(x==u) k
$<$%
- 1}}\\
\text{\texttt{\ \ \ \ else if(x%
$<$%
u) c(x,V,k
$<$%
- 1+f(x,V
$<$%
- V[V%
$<$%
u],k))}}\\
\text{\texttt{\ \ \ \ else c(x,V,k
$<$%
- 1+f(x,V
$<$%
- V[V%
$>$%
u],k)) }}\\
\text{\texttt{\ \ \ \ \}}}\\
\text{\texttt{\ \ k}}\\
\text{\texttt{\}}}%
\end{array}
\]

\noindent where $V$ is a random permutation on $\{1,3,5,\ldots,2n-1\}$ and $k$
is initially $0$. \ To model successful searches, let $x$ be a random odd
integer satisfying $1\leq x\leq2n-1$. \ To model unsuccessful searches, let
$x$ be a random even integer satisfying $0\leq x\leq2n$. \ This scenario is
exactly as described in \cite{Fi1-tcs2}. \ It is assumed, of course, that $x$
and $V$ are drawn independently with uniform sampling. \ We begin with even
$x$, because this case is simpler, followed by odd $x$. \ 

\subsection{Unsuccessful Search}

The probability generating function for $K_{n}$, given $n$, obeys a recursion
\cite{SF-tcs2}
\[%
\begin{array}
[c]{ccc}%
f_{n}(z)=\dfrac{2z+n-1}{n+1}f_{n-1}(z), &  & n\geq2;
\end{array}
\]%
\[
f_{1}(z)=z.
\]
Note that $f_{n}(1)=1$ always. \ Differentiating with respect to $z$:%
\[
f_{n}^{\prime}(z)=\dfrac{2}{n+1}f_{n-1}(z)+\dfrac{2z+n-1}{n+1}f_{n-1}^{\prime
}(z)
\]
we have first moment%
\[
\mathbb{E}(K_{n})=f_{n}^{\prime}(1)=\dfrac{2}{n+1}+f_{n-1}^{\prime}(1)
\]
that is,
\[
g_{n}=\dfrac{2}{n+1}+g_{n-1}%
\]
where $g_{k}=f_{k}^{\prime}(1)$ and $g_{1}=1$. \ Clearly $g_{2}=5/3$ and
$g_{3}=13/6$. \ Differentiating again:%
\[
f_{n}^{\prime\prime}(z)=\dfrac{4}{n+1}f_{n-1}^{\prime}(z)+\dfrac{2z+n-1}%
{n+1}f_{n-1}^{\prime\prime}(z)
\]
we have second factorial moment%
\[
\mathbb{E}(K_{n}(K_{n}-1))=f_{n}^{\prime\prime}(1)=\dfrac{4}{n+1}%
f_{n-1}^{\prime}(1)+f_{n-1}^{\prime\prime}(1),
\]
that is,%
\[
h_{n}=\dfrac{4}{n+1}g_{n-1}+h_{n-1}%
\]
where $h_{k}=f_{k}^{\prime\prime}(1)$ and $h_{1}=0$. \ Clearly $h_{2}=4/3$ and
$h_{3}=3$. \ Finally, we have variance%
\[
\mathbb{V}(K_{n})=h_{n}-g_{n}^{2}+g_{n}%
\]
which is $2/9$ when $n=2$ and $17/36$ when $n=3$. \ From (more typical)
harmonic number-based exact expressions, it can be proved that \cite{Mah-tcs2,
K1-tcs2, Ly-tcs2}%
\[
\mathbb{E}(K_{n})=2\ln(n)+2(\gamma-1)+\frac{3}{n}+o\left(  \frac{1}{n}\right)
,
\]%
\[
\mathbb{V}(K_{n})=2\ln(n)+2\left(  \gamma-\frac{\pi^{2}}{3}+1\right)
+\frac{7}{n}+o\left(  \frac{1}{n}\right)
\]
as $n\rightarrow\infty$.

\subsection{Successful Search}

The probability generating function for $K_{n}$, given $n$, obeys a recursion
\[%
\begin{array}
[c]{ccc}%
n^{2}f_{n}(z)=(n-1)(2z+n-1)f_{n-1}(z)+z, &  & n\geq2;
\end{array}
\]%
\[
f_{1}(z)=z.
\]
Note that $f_{n}(1)=1$ always. \ Differentiating with respect to $z$:%
\[
n^{2}f_{n}^{\prime}(z)=2(n-1)f_{n-1}(z)+(n-1)(2z+n-1)f_{n-1}^{\prime}(z)+1
\]
we have first moment%
\[
\mathbb{E}(K_{n})=f_{n}^{\prime}(1)=\frac{2(n-1)+(n-1)(n+1)f_{n-1}^{\prime
}(1)+1}{n^{2}}%
\]
that is,
\[
g_{n}=\frac{(2n-1)+\left(  n^{2}-1\right)  g_{n-1}}{n^{2}}%
\]
where $g_{k}=f_{k}^{\prime}(1)$ and $g_{1}=1$. \ Clearly $g_{2}=3/2$ and
$g_{3}=17/9$. \ Differentiating again:%
\[
n^{2}f_{n}^{\prime\prime}(z)=4(n-1)f_{n-1}^{\prime}(z)+(n-1)(2z+n-1)f_{n-1}%
^{\prime\prime}(z)
\]
we have second factorial moment%
\[
\mathbb{E}(K_{n}(K_{n}-1))=f_{n}^{\prime\prime}(1)=\frac{4(n-1)f_{n-1}%
^{\prime}(1)+(n-1)(n+1)f_{n-1}^{\prime\prime}(1)}{n^{2}},
\]
that is,%
\[
h_{n}=\frac{4(n-1)g_{n-1}+\left(  n^{2}-1\right)  h_{n-1}}{n^{2}}%
\]
where $h_{k}=f_{k}^{\prime\prime}(1)$ and $h_{1}=0$. \ Clearly $h_{2}=1$ and
$h_{3}=20/9$. \ Finally, we have variance $\mathbb{V}(K_{n})$ which is $1/4$
when $n=2$ and $44/81$ when $n=3$. \ 

It can be proved that \cite{Mah-tcs2, SF-tcs2, K1-tcs2, Ko-tcs2}%
\[
\mathbb{E}(K_{n})=2\ln(n)+(2\gamma-3)+\frac{2\ln(n)}{n}+\frac{2\gamma+1}%
{n}+o\left(  \frac{1}{n}\right)  ,
\]%
\begin{align*}
\mathbb{V}(K_{n})  &  =2\ln(n)+2\left(  \gamma-\frac{\pi^{2}}{3}+2\right)
-\frac{4\ln(n)^{2}}{n}+\frac{2(5-4\gamma)\ln(n)}{n}\\
&  +\left(  5+10\gamma-4\gamma^{2}-\frac{2\pi^{2}}{3}\right)  \frac{1}%
{n}+o\left(  \frac{1}{n}\right)
\end{align*}
as $n\rightarrow\infty$.

\subsection{Total Path Length}

The total (internal)\ path length $L_{n}$ is the sum of $K_{n}-1$ taken over
all odd integers $x$ from $1$ to $2n-1$. \ It is not surprising that
calculations are more involved here than before. The probability generating
function for $L_{n}$, given $n$, obeys a recursion \cite{K1-tcs2}%
\[%
\begin{array}
[c]{ccc}%
f_{n}(z)=\dfrac{z^{n-1}}{n}%
{\displaystyle\sum\limits_{k=0}^{n-1}}
f_{k}(z)f_{n-1-k}(z), &  & n\geq1;
\end{array}
\]%
\[
f_{0}(z)=1.
\]
Note that $f_{n}(1)=1$ always. \ Differentiating with respect to $z$:%
\[
f_{n}^{\prime}(z)=\dfrac{(n-1)z^{n-2}}{n}%
{\displaystyle\sum\limits_{k=0}^{n-1}}
f_{k}(z)f_{n-1-k}(z)+\dfrac{z^{n-1}}{n}%
{\displaystyle\sum\limits_{k=0}^{n-1}}
\left[  f_{k}^{\prime}(z)f_{n-1-k}(z)+f_{k}(z)f_{n-1-k}^{\prime}(z)\right]
\]
we have first moment%
\begin{align*}
\mathbb{E}(L_{n})  &  =f_{n}^{\prime}(1)=\dfrac{n-1}{n}\cdot n+\dfrac{1}{n}%
{\displaystyle\sum\limits_{k=0}^{n-1}}
\left[  f_{k}^{\prime}(1)+f_{n-1-k}^{\prime}(1)\right] \\
&  =n-1+\dfrac{2}{n}%
{\displaystyle\sum\limits_{k=0}^{n-1}}
f_{k}^{\prime}(1),
\end{align*}
that is,
\[
g_{n}=n-1+\dfrac{2}{n}%
{\displaystyle\sum\limits_{k=0}^{n-1}}
g_{k}%
\]
where $g_{k}=f_{k}^{\prime}(1)$ and $g_{0}=0$. \ Clearly $g_{1}=0$, $g_{2}=1$,
$g_{3}=8/3$ and $g_{4}=29/6$. \ Differentiating again:%
\begin{align*}
f_{n}^{\prime\prime}(z)  &  =\dfrac{(n-1)(n-2)z^{n-3}}{n}%
{\displaystyle\sum\limits_{k=0}^{n-1}}
f_{k}(z)f_{n-1-k}(z)+\dfrac{2(n-1)z^{n-2}}{n}%
{\displaystyle\sum\limits_{k=0}^{n-1}}
\left[  f_{k}^{\prime}(z)f_{n-1-k}(z)+f_{k}(z)f_{n-1-k}^{\prime}(z)\right] \\
&  +\dfrac{z^{n-1}}{n}%
{\displaystyle\sum\limits_{k=0}^{n-1}}
\left[  f_{k}^{\prime\prime}(z)f_{n-1-k}(z)+2f_{k}^{\prime}(z)f_{n-1-k}%
^{\prime}(z)+f_{k}(z)f_{n-1-k}^{\prime\prime}(z)\right]
\end{align*}
we have second factorial moment%
\begin{align*}
\mathbb{E}(L_{n}(L_{n}-1))  &  =f_{n}^{\prime\prime}(1)\\
&  =(n-1)(n-2)+2(n-1)\left[  f_{n}^{\prime}(1)-n+1\right]  +\dfrac{2}{n}%
{\displaystyle\sum\limits_{k=0}^{n-1}}
f_{k}^{\prime}(1)f_{n-1-k}^{\prime}(1)+\dfrac{2}{n}%
{\displaystyle\sum\limits_{k=0}^{n-1}}
f_{k}^{\prime\prime}(1),
\end{align*}
that is,%
\[
h_{n}=-(n-1)n+2(n-1)g_{n}+\dfrac{2}{n}%
{\displaystyle\sum\limits_{k=0}^{n-1}}
g_{k}g_{n-1-k}+\dfrac{2}{n}%
{\displaystyle\sum\limits_{k=0}^{n-1}}
h_{k}%
\]
where $h_{k}=f_{k}^{\prime\prime}(1)$ and $h_{0}=0$. \ Clearly $h_{1}=0$,
$h_{2}=0$, $h_{3}=14/3$ and $h_{4}=58/3$. \ Finally, we have variance
$\mathbb{V}(L_{n})$ which is $2/9$ when $n=3$ and $29/36$ when $n=4$.

It can be proved that \cite{Mah-tcs2, SF-tcs2, Wi-tcs2}
\[
\mathbb{E}(L_{n})=2n\ln(n)+2(\gamma-2)n+2\ln(n)+(2\gamma+1)+o(1),
\]%
\begin{align*}
\mathbb{V}(L_{n})  &  =\left(  7-\frac{2\pi^{2}}{3}\right)  n^{2}%
-2n\ln(n)+\left(  17-2\gamma-\frac{4\pi^{2}}{3}\right)  n\\
&  -2\ln(n)+\left(  5-2\gamma-\frac{2\pi^{2}}{3}\right)  +o(1)
\end{align*}
as $n\rightarrow\infty$.

\subsection{Higher Moments}

A\ third moment expression appears in \cite{MN-tcs2} for successful search;
analogous work for unsuccessful search remains undone. \ We focus on total
(internal)\ path length $L_{n}$ for BSTs. \ The cumulants $\kappa_{2}$,
$\kappa_{3}$, \ldots\ , $\kappa_{8}$ of $L_{n}$ were exhaustively studied by
Hennequin \cite{He1-tcs2, He2-tcs2}; these asymptotically satisfy%
\[
\kappa_{s}\sim\left[  a_{s}+(-1)^{s+1}2^{s}(s-1)!\zeta(s)\right]  n^{s}%
\]
as $n\rightarrow\infty$, where%
\[
\left\{  a_{s}\right\}  _{s=2}^{8}=\left\{  7,-19,\frac{937}{9},-\frac
{85981}{108},\frac{21096517}{2700},-\frac{7527245453}{81000},\frac
{19281922400989}{14883750}\right\}  .
\]
Hoffman \& Kuba \cite{HK-tcs2} obtained a complicated recurrence for an
associated sequence of rationals \cite{Cr-tcs2, EZ-tcs2}:%
\[
\left\{  c_{s}\right\}  _{s=2}^{8}=\left\{  7,-19,\frac{2260}{9}%
,-\frac{229621}{108},\frac{74250517}{2700},-\frac{30532750703}{81000}%
,\frac{90558126238639}{14883750}\right\}
\]
using what they called \textit{tiered binomial coefficients}. \ While they
utilized notation $(n,m)_{i}$, we adopt $T(i,n,m)$. \ It suffices to say that
$T(0,n,m)=\tbinom{n+m}{n}$ and a rich theory about $T(i,n,m)$ for $i\geq1$
awaits discovery. \ We give Mathematica code for generating $c_{s}$:%
\[%
\begin{array}
[c]{l}%
\text{\texttt{f[i\_,x\_,y\_] := (1/(i+1-x-y))
(Binomial[i-x,i]/Binomial[i-x-y,i])}}\\
\\
\text{\texttt{T[i\_,n\_,m\_] := If[n+m
$>$
0, Coefficient[Normal[\ \ }}\\
\text{\texttt{\ Series[f[i,x,y], \{x,0,n\}, \{y,0,m\}]], x\symbol{94}n
y\symbol{94}m], 1/(1+i)]}}\\
\\
\text{\texttt{c[s\_] := c[s] = ((s+1)/(s-1)) *}}\\
\text{\texttt{\ Sum[Sum[Sum[If[k1+k2+k3 == s, Multinomial[k1,k2,k3] c[k1]
c[k2] *}}\\
\text{\texttt{\ \ Sum[Sum[Sum[If[n+m+p == k3,}}\\
\text{\texttt{\ \ \ Sum[Multinomial[n,m,p] Binomial[m+k2,j] (-1)\symbol{94}j
(-2)\symbol{94}(n+m) n! m! T[n+k1+j,n,m],}}\\
\text{\texttt{\ \ \ \{j,0,m+k2\}], 0], }}\\
\text{\texttt{\ \ \{p,0,k3\}], \{m,0,k3\}], \{n,0,k3\}], 0], }}\\
\text{\texttt{\ \{k3,0,s\}], \{k2,0,s-1\}], \{k1,0,s-1\}] }}\\
\\
\text{\texttt{c[0] = 1;}}\\
\text{\texttt{c[1] = 0;}}%
\end{array}
\]
and code for generating $a_{s}$, given $c_{1}$, $c_{2}$, \ldots\ , $c_{s}$:%
\[%
\begin{array}
[c]{l}%
\text{\texttt{Sum[(-1)\symbol{94}(j-1) (j-1)! BellY[s, j, Table[c[i],
\{i,1,s-j+1\}]], \{j,1,s\}]}}%
\end{array}
\]
This final line employs a well-known expression for cumulants in terms of
partial (or incomplete) Bell polynomials of central moments. \ 

\section{Digital Search Trees}

Consider the R\ program:%
\[%
\begin{array}
[c]{l}%
\text{\texttt{f
$<$%
- function(x,M,p,k)}}\\
\text{\texttt{\{\ \ }}\\
\text{\texttt{\ \ q
$<$%
- NCOL(M)}}\\
\text{\texttt{\ \ if(NROW(M)==0) k
$<$%
- 0}}\\
\text{\texttt{\ \ else \{}}\\
\text{\texttt{\ \ \ \ if(all(x==matrix(M[1,],ncol=q))) k
$<$%
- 1}}\\
\text{\texttt{\ \ \ \ else \{}}\\
\text{\texttt{\ \ \ \ \ \ M
$<$%
- matrix(M[-1,],ncol=q)}}\\
\text{\texttt{\ \ \ \ \ \ M
$<$%
- matrix(M[M[,p]==x[p],],ncol=q) }}\\
\text{\texttt{\ \ \ \ \ \ k
$<$%
- 1+f(x,M,p
$<$%
- p+1,k)}}\\
\text{\texttt{\ \ \ \ \ \ \}}}\\
\text{\texttt{\ \ \ \ \}}}\\
\text{\texttt{\ \ k}}\\
\text{\texttt{\}}}%
\end{array}
\]
where $M$ is a random binary $n\times\ell$ matrix with $n$ distinct rows, $p$
is initially $1$ and $k$ is initially $0$. It is usually assumed
\cite{Mah-tcs2, K2-tcs2} that $\ell=\infty$, from which the row-distinctness
requirement follows almost surely (imagining the rows as binary expansions of
$n$ independent Uniform $[0,1]$ numbers). \ If instead $\ell=n$, as
exploratively specified in \cite{Fi2-tcs2}, then the matrix $M$ would need to
be generated carefully to avoid duplicate keys. To model successful searches,
let $x$ be a random row of $M$. \ To model unsuccessful searches, let $x$ be a
random binary $\ell$-vector that is not a row of $M$. \ 

\subsection{Unsuccessful Search}

The probability generating function for $K_{n}$, given $n$, is%
\[
\left\{
\begin{array}
[c]{ccc}%
\dfrac{1}{2}z+\dfrac{1}{2}z^{2} &  & \text{if }n=2,\\
\dfrac{1}{4}z+\dfrac{5}{8}z^{2}+\dfrac{1}{8}z^{3} &  & \text{if }n=3,\\
\dfrac{1}{8}z+\dfrac{19}{32}z^{2}+\dfrac{17}{64}z^{3}+\dfrac{1}{64}z^{4} &  &
\text{if }n=4,\\
\dfrac{1}{16}z+\dfrac{65}{128}z^{2}+\dfrac{195}{512}z^{3}+\dfrac{49}%
{1024}z^{4}+\dfrac{1}{1024}z^{5} &  & \text{if }n=5
\end{array}
\right.
\]
for $\ell=\infty$ and%
\[
\left\{
\begin{array}
[c]{ccc}%
\dfrac{2}{3}z+\dfrac{1}{3}z^{2} &  & \text{if }n=2,\\
\dfrac{2}{7}z+\dfrac{2}{3}z^{2}+\dfrac{1}{21}z^{3} &  & \text{if }n=3,\\
\dfrac{8}{65}z+\dfrac{302}{455}z^{2}+\dfrac{22}{105}z^{3}+\dfrac{1}{273}z^{4}
&  & \text{if }n=4,\\
\dfrac{52}{899}z+\dfrac{7384}{13485}z^{2}+\dfrac{34502}{94395}z^{3}+\dfrac
{26}{899}z^{4}+\dfrac{1}{6293}z^{5} &  & \text{if }n=5
\end{array}
\right.
\]
for $\ell=n$. \ A\ closed-form expression exists \cite{Mah-tcs2} for
$f_{n}(z)$ when $\ell=\infty$, but a corresponding simple recursive formula
does not evidently materialize. \ Section 3 contains verification of these
polynomial expressions.

\subsection{Successful Search}

The probability generating function for $K_{n}$, given $n$, is%
\[
\left\{
\begin{array}
[c]{ccc}%
\dfrac{1}{2}z+\dfrac{1}{2}z^{2} &  & \text{if }n=2,\\
\dfrac{1}{3}z+\dfrac{1}{2}z^{2}+\dfrac{1}{6}z^{3} &  & \text{if }n=3,\\
\dfrac{1}{4}z+\dfrac{7}{16}z^{2}+\dfrac{9}{32}z^{3}+\dfrac{1}{32}z^{4} &  &
\text{if }n=4,\\
\dfrac{1}{5}z+\dfrac{3}{8}z^{2}+\dfrac{11}{32}z^{3}+\dfrac{5}{64}z^{4}%
+\dfrac{1}{320}z^{5} &  & \text{if }n=5
\end{array}
\right.
\]
for $\ell=\infty$ and%
\[
\left\{
\begin{array}
[c]{ccc}%
\dfrac{1}{2}z+\dfrac{1}{2}z^{2} &  & \text{if }n=2,\\
\dfrac{1}{3}z+\dfrac{11}{21}z^{2}+\dfrac{1}{7}z^{3} &  & \text{if }n=3,\\
\dfrac{1}{4}z+\dfrac{9}{20}z^{2}+\dfrac{39}{140}z^{3}+\dfrac{3}{140}z^{4} &  &
\text{if }n=4,\\
\dfrac{1}{5}z+\dfrac{1707}{4495}z^{2}+\dfrac{23561}{67425}z^{3}+\dfrac
{4657}{67425}z^{4}+\dfrac{39}{22475}z^{5} &  & \text{if }n=5
\end{array}
\right.
\]
for $\ell=n$. \ A\ closed-form expression exists \cite{Lou-tcs2, Mah-tcs2} for
$f_{n}(z)$ when $\ell=\infty$, but a corresponding simple recursive formula
again does not materialize. \ Means and variances for $\ell=\infty$ and those
for $\ell=n$ unsurprisingly become closer as $n$ increases.

\subsection{Total Path Length}

The total (internal)\ path length $L_{n}$ is the sum of $K_{n}-1$ taken over
all rows $m$ of $M$. \ It is not surprising that calculations are more
involved here than before. \ Assume that $\ell=\infty$. The probability
generating function for $L_{n}$, given $n$, obeys a recursion \cite{KPS-tcs2}%
\[%
\begin{array}
[c]{ccc}%
f_{n}(z)=z^{n-1}2^{1-n}%
{\displaystyle\sum\limits_{k=0}^{n-1}}
\dbinom{n-1}{k}f_{k}(z)f_{n-1-k}(z), &  & n\geq1;
\end{array}
\]%
\[
f_{0}(z)=1.
\]
Note that $f_{n}(1)=1$ always. \ Differentiating with respect to $z$:%
\begin{align*}
f_{n}^{\prime}(z)  &  =(n-1)z^{n-2}2^{1-n}%
{\displaystyle\sum\limits_{k=0}^{n-1}}
\dbinom{n-1}{k}f_{k}(z)f_{n-1-k}(z)\\
&  +z^{n-1}2^{1-n}%
{\displaystyle\sum\limits_{k=0}^{n-1}}
\dbinom{n-1}{k}\left[  f_{k}^{\prime}(z)f_{n-1-k}(z)+f_{k}(z)f_{n-1-k}%
^{\prime}(z)\right]
\end{align*}
we have first moment%
\[
\mathbb{E}(L_{n})=f_{n}^{\prime}(1)=n-1+2^{2-n}%
{\displaystyle\sum\limits_{k=0}^{n-1}}
\dbinom{n-1}{k}f_{k}^{\prime}(1)
\]
that is,
\[
g_{n}=n-1+2^{2-n}%
{\displaystyle\sum\limits_{k=0}^{n-1}}
\dbinom{n-1}{k}g_{k}%
\]
where $g_{k}=f_{k}^{\prime}(1)$ and $g_{0}=0$. \ Clearly $g_{1}=0$, $g_{2}=1$,
$g_{3}=5/2$ and $g_{4}=35/8$. \ Differentiating again:%
\begin{align*}
f_{n}^{\prime\prime}(z)  &  =(n-1)(n-2)z^{n-3}2^{1-n}%
{\displaystyle\sum\limits_{k=0}^{n-1}}
\dbinom{n-1}{k}f_{k}(z)f_{n-1-k}(z)\\
&  +(n-1)z^{n-2}2^{2-n}%
{\displaystyle\sum\limits_{k=0}^{n-1}}
\dbinom{n-1}{k}\left[  f_{k}^{\prime}(z)f_{n-1-k}(z)+f_{k}(z)f_{n-1-k}%
^{\prime}(z)\right] \\
&  +z^{n-1}2^{1-n}%
{\displaystyle\sum\limits_{k=0}^{n-1}}
\dbinom{n-1}{k}\left[  f_{k}^{\prime\prime}(z)f_{n-1-k}(z)+2f_{k}^{\prime
}(z)f_{n-1-k}^{\prime}(z)+f_{k}(z)f_{n-1-k}^{\prime\prime}(z)\right]
\end{align*}
we have second factorial moment%
\begin{align*}
\mathbb{E}(L_{n}(L_{n}-1))  &  =f_{n}^{\prime\prime}(1)\\
&  =(n-1)(n-2)+2(n-1)\left[  f_{n}^{\prime}(1)-n+1\right] \\
&  +2^{2-n}%
{\displaystyle\sum\limits_{k=0}^{n-1}}
\dbinom{n-1}{k}f_{k}^{\prime}(1)f_{n-1-k}^{\prime}(1)+2^{2-n}%
{\displaystyle\sum\limits_{k=0}^{n-1}}
\dbinom{n-1}{k}f_{k}^{\prime\prime}(1),
\end{align*}
that is,%
\[
h_{n}=-(n-1)n+2(n-1)g_{n}+2^{2-n}%
{\displaystyle\sum\limits_{k=0}^{n-1}}
\dbinom{n-1}{k}g_{k}g_{n-1-k}+2^{2-n}%
{\displaystyle\sum\limits_{k=0}^{n-1}}
\dbinom{n-1}{k}h_{k}%
\]
where $h_{k}=f_{k}^{\prime\prime}(1)$ and $h_{0}=0$. \ Clearly $h_{1}=0$,
$h_{2}=0$, $h_{3}=4$ and $h_{4}=61/4$. \ Finally, we have variance
$\mathbb{V}(L_{n})$ which is $1/4$ when $n=3$ and $31/64$ when $n=4$.

Define constants%
\[%
\begin{array}
[c]{ccccc}%
\alpha=%
{\displaystyle\sum\limits_{j=1}^{\infty}}
\dfrac{1}{2^{j}-1}, &  & \beta=%
{\displaystyle\sum\limits_{j=1}^{\infty}}
\dfrac{1}{\left(  2^{j}-1\right)  ^{2}}, &  & Q=%
{\displaystyle\prod\limits_{j=1}^{\infty}}
\left(  1-\dfrac{1}{2^{j}}\right)  .
\end{array}
\]
Let $Q_{l}$ denote the $l^{\text{th}}$ partial product of $Q$ and%
\[
\varphi(x)=\left\{
\begin{array}
[c]{lll}%
\dfrac{x-\ln(x)-1}{(x-1)^{2}} &  & \text{if }x\neq1,\\
\dfrac{1}{2} &  & \text{if }x=1.
\end{array}
\right.
\]
It can be proved that \cite{KPS-tcs2, HFZ-tcs2}%
\begin{align*}
\mathbb{E}(L_{n}) &  =\frac{n\ln(n)}{\ln(2)}+n\left(  \frac{\gamma-1}{\ln
(2)}+\frac{1}{2}-\alpha+\delta_{1}\left(  n\right)  \right)  +\frac{\ln
(n)}{\ln(2)}\\
&  +\left(  \frac{2\gamma-1}{2\ln(2)}+\frac{5}{2}-\alpha\right)  +\delta
_{2}\left(  n\right)  +O\left(  \frac{\ln(n)}{n}\right)  ,
\end{align*}%
\[
\mathbb{V}(L_{n})=n\left(  C+\delta_{3}\left(  n\right)  \right)  +O\left(
\frac{\ln(n)^{2}}{n}\right)
\]
as $n\rightarrow\infty$, where
\[
C=\frac{Q}{\ln(2)}%
{\displaystyle\sum\limits_{j,k,l\geq0}}
\frac{(-1)^{j}}{Q_{j}Q_{k}Q_{l}}2^{-j(j+1)/2-k-l}\varphi\left(  2^{-j-k}%
+2^{-j-l}\right)  =0.2660036454....
\]
This expression for $C$ is, needless to say, a stunning result.

Assuming instead that $\ell=n$, all we currently possess are PGFs for small
$n$:
\[
\left\{
\begin{array}
[c]{ccc}%
z &  & \text{if }n=2,\\
\dfrac{4}{7}z^{2}+\dfrac{3}{7}z^{3} &  & \text{if }n=3,\\
\dfrac{4}{5}z^{4}+\dfrac{4}{35}z^{5}+\dfrac{3}{35}z^{6} &  & \text{if }n=4,\\
\dfrac{8984}{13485}z^{6}+\dfrac{3136}{13485}z^{7}+\dfrac{364}{4495}%
z^{8}+\dfrac{52}{4495}z^{9}+\dfrac{39}{4495}z^{10} &  & \text{if }n=5.
\end{array}
\right.
\]
A deeper understanding of finite-key DSTs would be welcome.

\subsection{Some Combinatorics}

We focus on unsuccessful searches, for both infinite keys ($\ell=\infty$) and
finite keys ($\ell=n$). \ Let us examine the coefficients of $z^{n}$ and $z$
for simplicity. \ The digital search trees appearing in Figure 1 for $n=2$
proceed from matrices%
\[
\left(
\begin{array}
[c]{c}%
M\\
x
\end{array}
\right)  =\left(
\begin{array}
[c]{cc}%
a & b\\
1 & c\\
1 & d
\end{array}
\right)  ,\;\;\left(
\begin{array}
[c]{cc}%
a & b\\
0 & c\\
0 & d
\end{array}
\right)  ,\;\;\left(
\begin{array}
[c]{cc}%
a & b\\
0 & c\\
1 & d
\end{array}
\right)  ,\;\;\left(
\begin{array}
[c]{cc}%
a & b\\
1 & c\\
0 & d
\end{array}
\right)
\]
respectively. \ When $\ell=\infty$, the indicated keys are merely
abbreviations (two leading bits in an infinite sequence); hence the keys are
automatically distinct; thus%
\[
\mathbb{P}\left\{  K_{2}=2\right\}  =\frac{2\cdot2^{4}}{2^{6}}=\frac{1}{2},
\]%
\[
\mathbb{P}\left\{  K_{2}=1\right\}  =\frac{2\cdot2^{4}}{2^{6}}=\frac{1}{2}%
\]
where $2^{(n+1)n}$ is the count of $(n+1)\times n$ binary matrices. \ When
$\ell=n$, however, key-distinctness must be manually enforced. \ We obtain the
condition%
\[
(a,b)\neq(1,c),\;\;(a,b)\neq(1,d)\;\;\&\;\;(1,c)\neq(1,d)
\]
which is equivalent to $a=0\;\;\&\;\;d=1-c$ and gives $4$ possibilities; also
the condition%
\[
(a,b)\neq(0,c),\;\;(a,b)\neq(1,d)\;\;\&\;\;(0,c)\neq(1,d)
\]
which is equivalent to $(1-a)c+ad=1-b$ and gives $8$ possibilities; therefore%
\[
\mathbb{P}\left\{  \tilde{K}_{2}=2\right\}  =\frac{2\cdot4}{4!/1!}=\frac{1}%
{3},
\]%
\[
\mathbb{P}\left\{  \tilde{K}_{2}=1\right\}  =\frac{2\cdot8}{4!/1!}=\frac{2}{3}%
\]
where $\left(  2^{n}\right)  !/\left(  2^{n}-n-1\right)  !$ is the count of
permutations of $2^{n}$ objects, taken $n+1$ at a time. \
\begin{figure}
[ptb]
\begin{center}
\includegraphics[
height=2.0185in,
width=6.6115in
]%
{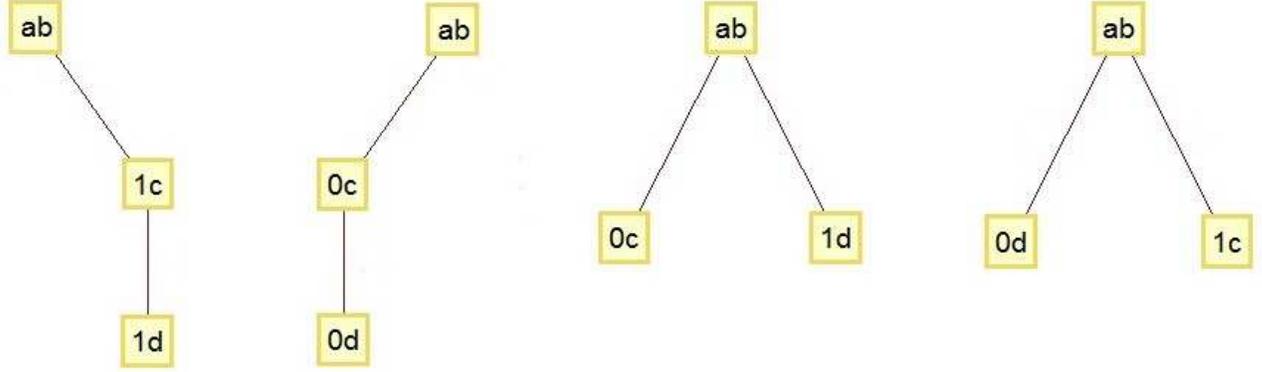}%
\caption{Two linear cases and two triangular cases for $n=2$.}%
\end{center}
\end{figure}

For $n=3$ and $\ell=\infty$, using Figures 2 and 3, we have%
\[
\mathbb{P}\left\{  K_{3}=3\right\}  =\frac{4\cdot2^{7}}{2^{12}}=\frac{1}{8},
\]%
\[
\mathbb{P}\left\{  K_{3}=1\right\}  =\frac{2\cdot2^{9}}{2^{12}}=\frac{1}{4}%
\]
but when $\ell=n$ instead, we have%
\[
\mathbb{P}\left\{  \tilde{K}_{3}=3\right\}  =\frac{4\cdot20}{8!/4!}=\frac
{1}{21},
\]%
\[
\mathbb{P}\left\{  \tilde{K}_{3}=1\right\}  =\frac{2\cdot240}{8!/4!}=\frac
{2}{7}.
\]%
\begin{figure}
[ptb]
\begin{center}
\includegraphics[
height=2.8522in,
width=6.608in
]%
{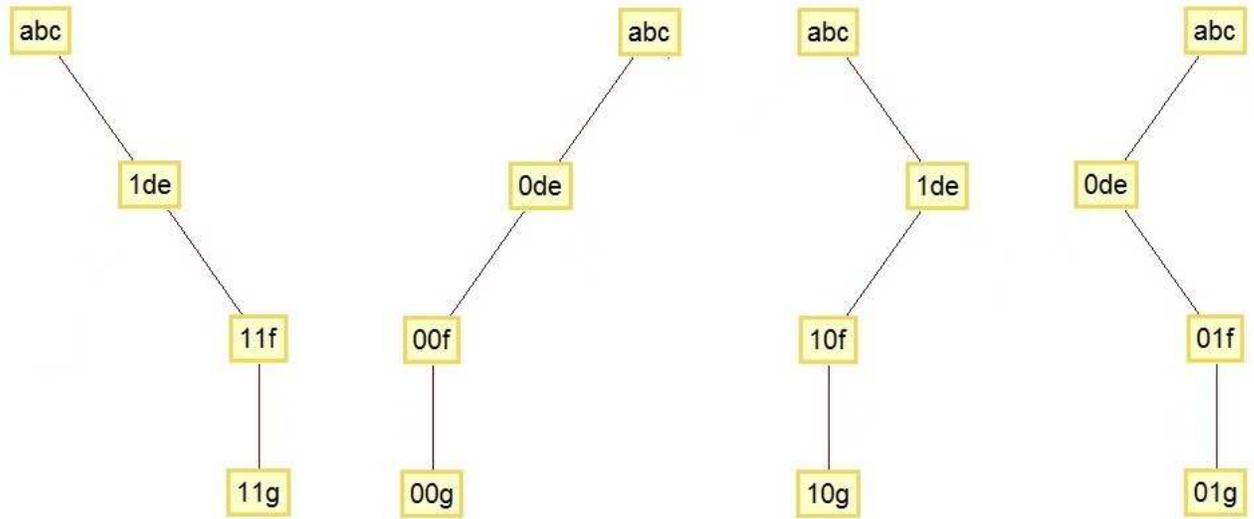}%
\caption{Four linear cases for $n=3$; note that two are reflections of the
others.}%
\end{center}
\end{figure}
\begin{figure}
[ptb]
\begin{center}
\includegraphics[
height=3.0242in,
width=5.6723in
]%
{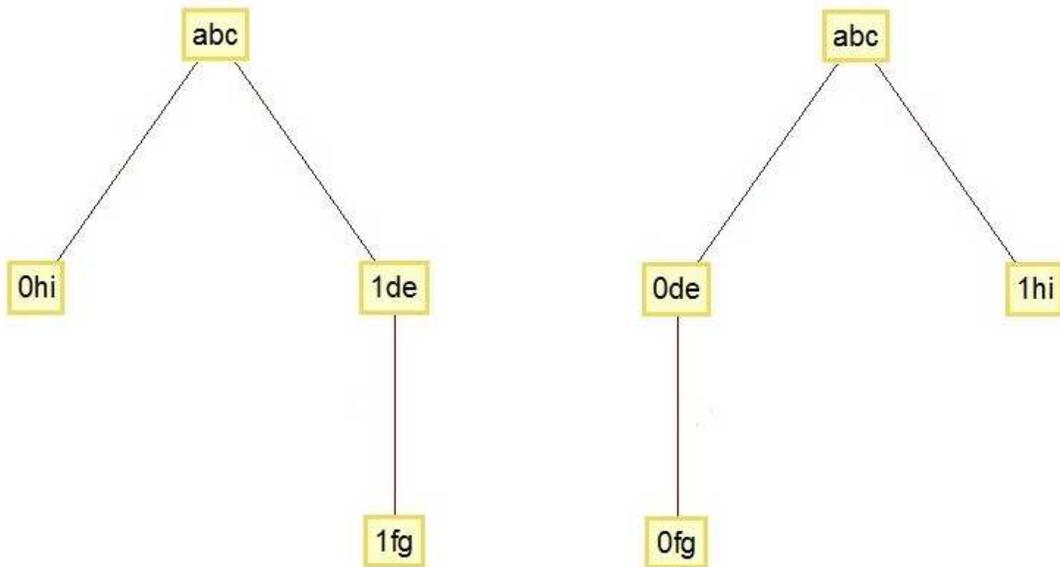}%
\caption{Two triangular cases for $n=3$; note that one is a reflection of the
other.}%
\end{center}
\end{figure}

For $n=4$ and $\ell=\infty$, using Figures 4, 5 and 6, we have%
\[
\mathbb{P}\left\{  K_{4}=4\right\}  =\frac{8\cdot2^{11}}{2^{20}}=\frac{1}%
{64},
\]%
\[
\mathbb{P}\left\{  K_{4}=1\right\}  =\frac{4\cdot2^{14}+4\cdot2^{14}}{2^{20}%
}=\frac{1}{8}%
\]
but when $\ell=n$ instead, we have%
\[
\mathbb{P}\left\{  \tilde{K}_{4}=4\right\}  =\frac{8\cdot240}{16!/11!}%
=\frac{1}{273},
\]%
\[
\mathbb{P}\left\{  \tilde{K}_{4}=1\right\}  =\frac{4\cdot6912+4\cdot
9216}{16!/11!}=\frac{8}{65}.
\]
The emergence of bi-triangular cases at $n=4$ complicates our study for
$n\geq5$. \ A similar argument for coefficients of $z^{2}$, \ldots\ ,
$z^{n-1}$, as well as for successful searches, is possible.%
\begin{figure}
[ptb]
\begin{center}
\includegraphics[
height=2.597in,
width=6.608in
]%
{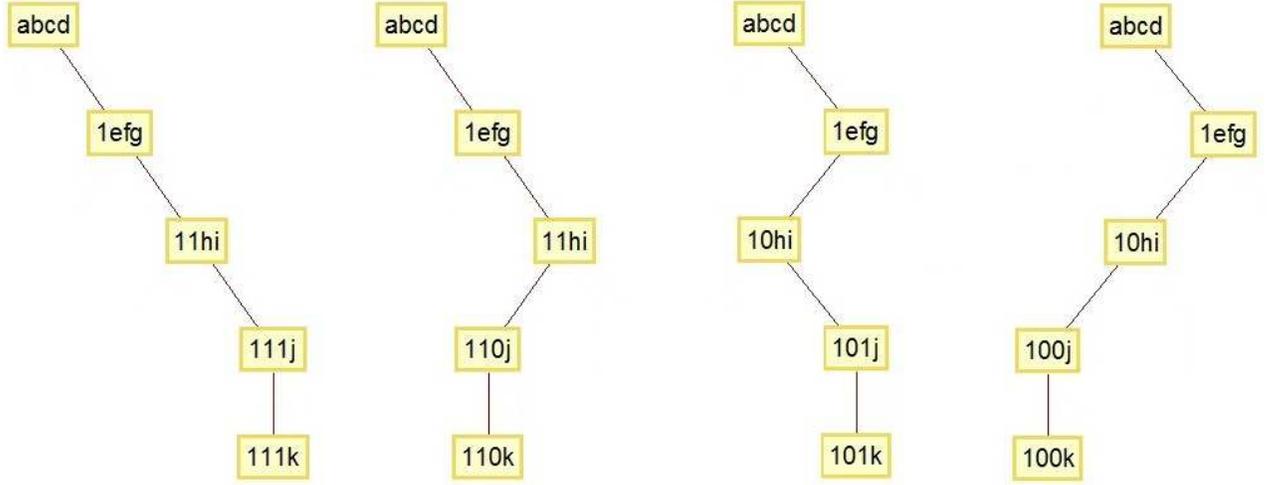}%
\caption{Eight linear cases for $n=4$ (these four cases plus their
reflections).}%
\end{center}
\end{figure}
\begin{figure}
[ptb]
\begin{center}
\includegraphics[
height=3.0242in,
width=5.3731in
]%
{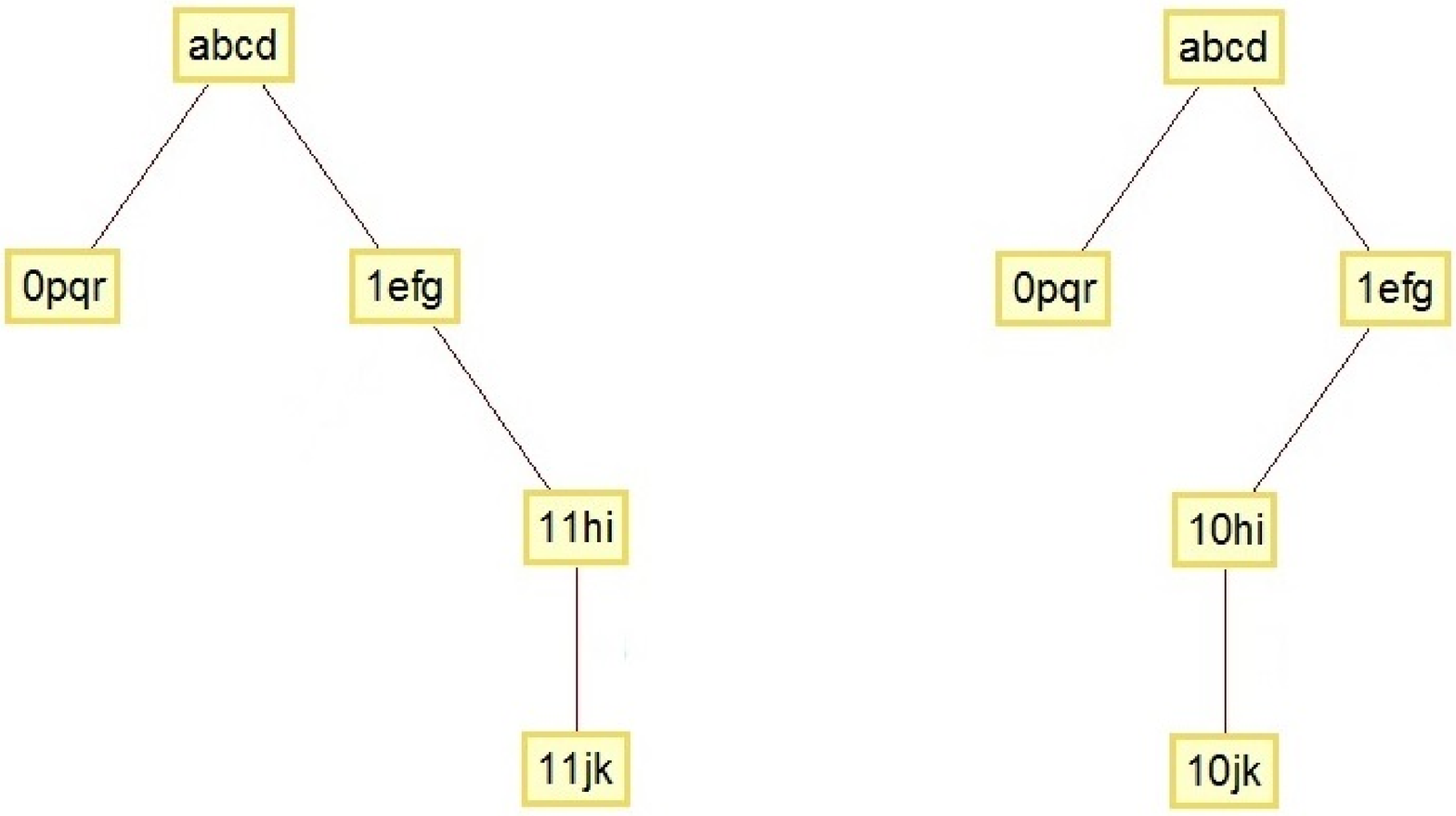}%
\caption{Four triangular cases for $n=4$ (these two cases plus their
reflections).}%
\end{center}
\end{figure}
\begin{figure}
[ptb]
\begin{center}
\includegraphics[
height=3.0242in,
width=6.0917in
]%
{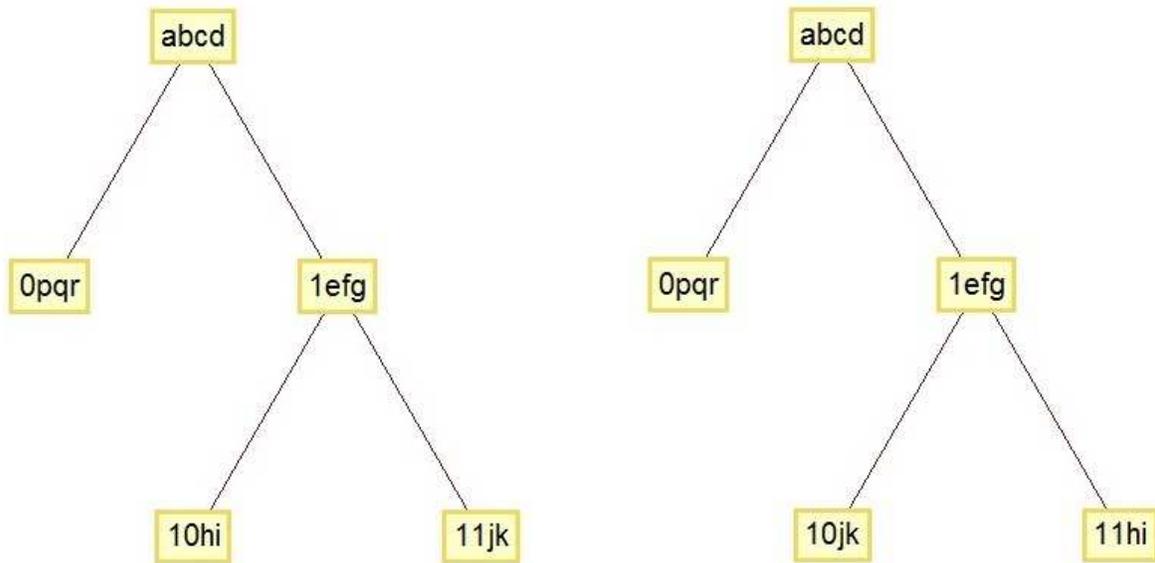}%
\caption{Four bi-triangular cases for $n=4$ (these two cases plus their
reflections).}%
\end{center}
\end{figure}

Third and fourth moment expressions appear in \cite{LP-tcs2} for unsuccessful
search on infinite keys. \ The covariance between two random distinct
successful search costs within the same tree is apparently $\sim D/n$ as
$n\rightarrow\infty$, where \cite{KPS-tcs2}
\[
D=C-\frac{1}{12}-\frac{\pi^{2}}{6\ln(2)^{2}}+\alpha+\beta=-0.4970105417....
\]
Verifying this interesting result via simulation remains open. \ What can be
said about the cost covariance for two distinct unsuccessful searches? \ What
can be said about the cost covariance given a successful search and an
unsuccessful search? \ 

\section{Acknowledgements}

I\ am grateful to Markus Kuba and Sumit Kumar Jha for helpful discussions, and
to David Penman for providing \cite{He2-tcs2} (which at one time was available
at http://algo.inria.fr/).

\end{document}